\documentclass{svproc}
\usepackage{url}

\usepackage{graphicx}
\usepackage{float}

\usepackage{helvet}         
\usepackage{courier}        
\usepackage{type1cm}        
\usepackage{makeidx}         
\usepackage{graphicx}        
\usepackage{multicol}        
\usepackage[bottom]{footmisc}

\usepackage[all,cmtip]{xy}

\usepackage{amsfonts}
\usepackage{mathrsfs}
\usepackage{amscd}
\usepackage{amsmath}
\usepackage{amssymb}
\usepackage{latexsym}
\usepackage{lscape}
\usepackage{comment}
\usepackage{amscd}
\usepackage{wasysym}  
\usepackage{tikz}
\usetikzlibrary{matrix,arrows}
\usepackage{tikz-cd}
\usepackage{appendix}
\usepackage[nice]{nicefrac}
\usepackage{dsfont, stmaryrd}
\usepackage{xparse}
\usepackage{listings}
\usepackage{lstautogobble}
\usepackage{lstlinebgrd}
\usepackage{bbm}
\usepackage[scaled]{inconsolata}
\usepackage{xcolor}
\usepackage{caption}
\definecolor{keywordcolor}{rgb}{0.7, 0.1, 0.1}   
\definecolor{tacticcolor}{rgb}{0.0, 0.1, 0.6}    
\definecolor{commentcolor}{rgb}{0.4, 0.4, 0.4}   
\definecolor{symbolcolor}{rgb}{0.0, 0.1, 0.6}    
\definecolor{sortcolor}{rgb}{0.1, 0.5, 0.1}      
\definecolor{attributecolor}{rgb}{0.7, 0.1, 0.1} 

\usetikzlibrary{decorations.markings}

\makeatletter
\tikzcdset{
  open/.code     = {\tikzcdset{hook, circled};},
  closed/.code   = {\tikzcdset{hook, slashed};},
  open'/.code    = {\tikzcdset{hook', circled};},
  closed'/.code  = {\tikzcdset{hook', slashed};},
  circled/.code  = {\tikzcdset{markwith = {\draw (0,0) circle (.375ex);}};},
  slashed/.code  = {\tikzcdset{markwith = {\draw[-] (-.4ex,-.4ex) -- (.4ex,.4ex);}};},
  markwith/.code ={
    \pgfutil@ifundefined%
    {tikz@library@decorations.markings@loaded}%
    {\pgfutil@packageerror{tikz-cd}{You need to say %
      \string\usetikzlibrary{decorations.markings} to use arrows with markings}{}}{}%
    \pgfkeysalso{/tikz/postaction = {
      /tikz/decorate,
      /tikz/decoration={markings, mark = at position 0.5 with {#1}}}
    }
  },
}
\makeatother

\renewcommand{\geq}{\geqslant}
\renewcommand{\leq}{\leqslant}

\newtheorem{thm}{Theorem}[section]

\newtheorem{propo}[thm]{Proposition}

\newtheorem{propodef}[thm]{Definition and Proposition}
\newtheorem{lem}[thm]{Lemma}
\newtheorem{sublem}[thm]{Sublemma}
\newtheorem{lem-def}[thm]{Lemma-Definition}
\newtheorem{cor}[thm]{Corollary}
\newtheorem{conject}[thm]{Conjecture}
\newtheorem{propert}[thm]{Properties}
\newtheorem{observ}[thm]{Observation}

\newtheorem{fac}[thm]{Fact}

\newtheorem{notat}[thm]{Notation}

\newtheorem{ex}[thm]{Example}

\newtheorem{rmk}[thm]{Remark}
\newtheorem{dfn}[thm]{Definition}
\newtheorem{quest}[thm]{Question}

\numberwithin{equation}{section}

\newcommand{\nc}{\newcommand}

\nc{\abst}{\begin{abs}} \nc{\xabst}{\end{abs}}
\nc{\theo}{\begin{thm}} \nc{\xtheo}{\end{thm}}
\nc{\prop}{\begin{propo}} \nc{\xprop}{\end{propo}}

\nc{\nota}{\begin{notat}} \nc{\xnota}{\end{notat}}
\nc{\depr}{\begin{propodef}} \nc{\xdepr}{\end{propodef}}
\nc{\lemm}{\begin{lem}} \nc{\xlemm}{\end{lem}}
\nc{\sublemm}{\begin{sublem}} \nc{\xsublemm}{\end{sublem}}
\nc{\lemmdefi}{\begin{lem-def}} \nc{\xlemmdefi}{\end{lem-def}}
\nc{\coro}{\begin{cor}} \nc{\xcoro}{\end{cor}}
\nc{\conj}{\begin{conject}} \nc{\xconj}{\end{conject}}
\nc{\proper}{\begin{propert}} \nc{\xproper}{\end{propert}}
\nc{\obse}{\begin{observ}} \nc{\xobse}{\end{observ}}
\nc{\ques}{\begin{quest}} \nc{\xques}{\end{quest}}

\nc{\fact}{\begin{fac}} \nc{\xfact}{\end{fac}}
\nc{\ackn}{\begin{ack}} \nc{\xackn}{\end{ack}}
\nc{\exam}{\begin{ex}} \nc{\xexam}{\end{ex}}
\nc{\rema}{\begin{rmk}} \nc{\xrema}{\end{rmk}}
\nc{\defi}{\begin{dfn}} \nc{\xdefi}{\end{dfn}}

\nc{\pf}{\begin{proof}} \nc{\xpf}{\end{proof}}

\nc{\nn}{\mathrm{n}}
\nc{\on}{\operatorname}
\nc{\fraka}{{\mathfrak a}} \nc{\bba}{{\mathbf a}}
\nc{\frakb}{{\mathfrak b}}
\nc{\frakc}{{\mathfrak c}}
\nc{\frakd}{{\mathfrak d}}
\nc{\frake}{{\mathfrak e}}
\nc{\frakf}{{\mathfrak f}}
\nc{\frakg}{{\mathfrak g}}
\nc{\frakh}{{\mathfrak h}}
\nc{\fraki}{{\mathfrak i}}
\nc{\frakj}{{\mathfrak j}}
\nc{\frakk}{{\mathfrak k}}
\nc{\frakl}{{\mathfrak l}}
\nc{\frakm}{{\mathfrak m}}
\nc{\frakn}{{\mathfrak n}}
\nc{\frako}{{\mathfrak o}}
\nc{\frakp}{{\mathfrak p}}
\nc{\frakq}{{\mathfrak q}}
\nc{\frakr}{{\mathfrak r}}
\nc{\fraks}{{\mathfrak s}}
\nc{\frakt}{{\mathfrak t}}
\nc{\fraku}{{\mathfrak u}}
\nc{\frakv}{{\mathfrak v}}
\nc{\frakw}{{\mathfrak w}}
\nc{\frakx}{{\mathfrak x}}
\nc{\fraky}{{\mathfrak y}}
\nc{\frakz}{{\mathfrak z}}
\nc{\frakA}{{\mathfrak A}}
\nc{\frakB}{{\mathfrak B}}
\nc{\frakC}{{\mathfrak C}}
\nc{\frakD}{{\mathfrak D}}
\nc{\frakE}{{\mathfrak E}}
\nc{\frakF}{{\mathfrak F}}
\nc{\frakG}{{\mathfrak G}}
\nc{\frakH}{{\mathfrak H}}
\nc{\frakI}{{\mathfrak I}}
\nc{\frakJ}{{\mathfrak J}}
\nc{\frakK}{{\mathfrak K}}
\nc{\frakL}{{\mathfrak L}}
\nc{\frakM}{{\mathfrak M}}
\nc{\frakN}{{\mathfrak N}}
\nc{\frakO}{{\mathfrak O}}
\nc{\frakP}{{\mathfrak P}}
\nc{\frakQ}{{\mathfrak Q}}
\nc{\frakR}{{\mathfrak R}}
\nc{\frakS}{{\mathfrak S}}
\nc{\frakT}{{\mathfrak T}}
\nc{\frakU}{{\mathfrak U}}
\nc{\frakV}{{\mathfrak V}}
\nc{\frakW}{{\mathfrak W}}
\nc{\frakX}{{\mathfrak X}}
\nc{\frakY}{{\mathfrak Y}}
\nc{\frakZ}{{\mathfrak Z}}
\nc{\bbA}{{\mathbb A}}
\nc{\bbB}{{\mathbb B}}
\nc{\bbC}{{\mathbb C}}
\nc{\bbD}{{\mathbb D}}
\nc{\bbE}{{\mathbb E}}
\nc{\bbF}{{\mathbb F}} \nc{\bbf}{{\mathbf f}}
\nc{\bbG}{{\mathbb G}}
\nc{\bbH}{{\mathbb H}}
\nc{\bbI}{{\mathbb I}}
\nc{\bbJ}{{\mathbb J}}
\nc{\bbK}{{\mathbb K}}
\nc{\bbL}{{\mathbb L}}
\nc{\bbM}{{\mathbb M}}
\nc{\bbN}{{\mathbb N}}
\nc{\bbO}{{\mathbb O}}
\nc{\bbP}{{\mathbb P}}
\nc{\bbQ}{{\mathbb Q}}
\nc{\bbR}{{\mathbb R}}
\nc{\bbS}{{\mathbb S}}
\nc{\bbT}{{\mathbb T}}
\nc{\bbU}{{\mathbb U}}
\nc{\bbV}{{\mathbb V}}
\nc{\bbW}{{\mathbb W}}
\nc{\bbX}{{\mathbb X}}
\nc{\bbY}{{\mathbb Y}}
\nc{\bbZ}{{\mathbb Z}}
\nc{\calA}{{\mathcal A}}
\nc{\calB}{{\mathcal B}}
\nc{\calC}{{\mathcal C}}
\nc{\calD}{{\mathcal D}}
\nc{\calE}{{\mathcal E}}
\nc{\calF}{{\mathcal F}}
\nc{\calG}{{\mathcal G}}
\nc{\calH}{{\mathcal H}}
\nc{\calI}{{\mathcal I}}
\nc{\calJ}{{\mathcal J}}
\nc{\calK}{{\mathcal K}}
\nc{\calL}{{\mathcal L}}
\nc{\calM}{{\mathcal M}}
\nc{\calN}{{\mathcal N}}
\nc{\calO}{{\mathcal O}}
\nc{\calP}{{\mathcal P}}
\nc{\calQ}{{\mathcal Q}}
\nc{\calR}{{\mathcal R}}
\nc{\calS}{{\mathcal S}}
\nc{\calT}{{\mathcal T}}
\nc{\calU}{{\mathcal U}}
\nc{\calV}{{\mathcal V}}
\nc{\calW}{{\mathcal W}}
\nc{\calX}{{\mathcal X}}
\nc{\calY}{{\mathcal Y}}
\nc{\calZ}{{\mathcal Z}}

\nc{\Spac}{{\mathrm{Spaces}}}
\nc{\scrA}{{\mathscr A}}
\nc{\scrE}{{\mathscr E}}
\nc{\scrR}{{\mathscr R}}

\nc{\Bmu}{\mbox{$\raisebox{-0.59ex}{$l$}\hspace{-0.18em}\mu\hspace{-0.88em}\raisebox{-0.98ex}{\scalebox{2}{$\color{white}.$}}\hspace{-0.416em}\raisebox{+0.88ex}{$\color{white}.$}\hspace{0.46em}$}{}}

\nc{\bnu}{{\bar{ \nu}}}

\nc{\olO}{\bar{\calO}}

\nc{\al}{{\alpha}} 
\nc{\be}{{\beta}}
\nc{\ga}{{\gamma}} \nc{\Ga}{{\Gamma}}
 \nc{\hGa}{\hat{\Gamma}}
\nc{\ve}{{\varepsilon}} 
\nc{\la}{{\lambda}} \nc{\La}{{\Lambda}}
\nc{\om}{\omega} \nc{\Om}{\Omega} 
\nc{\sig}{{\sigma}} \nc{\Sig}{{\Sigma}}

\nc{\tnb}{\psi_{\rm tame}}
\nc{\oM}{\overline{{M}}}
\nc{\op}{{\on{op}}}
\nc{\ad}{{\on{ad}}}
\nc{\alg}{{\on{alg}}}
\nc{\Ad}{{\on{Ad}}}
\nc{\Adm}{{\on{Adm}}} \nc{\aff}{{\on{aff}}}
\nc{\Aut}{{\on{Aut}}}
\nc{\Bun}{{\on{Bun}}}
\nc{\cha}{{\on{char}}}
\nc{\der}{{\on{der}}}
\nc{\Der}{{\on{Der}}}
\nc{\diag}{{\on{diag}}}
\nc{\End}{{\on{End}}}
\nc{\Fl}{{\calF\!\ell}}
\nc{\Tr}{{\on{Transp}}}
\nc{\TR}{{\calT\!\calR}}
\nc{\Gal}{{\on{Gal}}}
\nc{\Gr}{{\on{Gr}}}
\nc{\rH}{{\on{H}}}
\nc{\Hom}{{\on{Hom}}}
\nc{\IC}{{\on{IC}}}
\nc{\id}{{\on{id}}}
\nc{\Id}{{\on{Id}}}
\nc{\ind}{{\on{ind}}}
\nc{\Ind}{{\on{Ind}}}
\nc{\Lie}{{\on{Lie}}}
\nc{\Pic}{{\on{Pic}}}
\nc{\pr}{{\on{pr}}}
\nc{\Res}{{\on{Res}}}
\nc{\res}{{\on{res}}} \nc{\Sat}{{\on{Sat}}}
\nc{\s}{{\on{sc}}}
\nc{\drv}{{\on{der}}}
\nc{\sgn}{{\on{sgn}}}
\nc{\Spec}{{\on{Spec}}}\nc{\Spf}{\on{Spf}} 
\nc{\Sph}{\on{Sph}}
\nc{\St}{{\on{St}}}
\nc{\tr}{{\on{tr}}}
\nc{\Mod}{{\mathrm{-Mod}}}
\nc{\Hilb}{{\on{Hilb}}} 
\nc{\Ext}{{\on{Ext}}} 
\nc{\vs}{{\on{Vec}}}
\nc{\ev}{{\on{ev}}}
\nc{\nO}{{\breve{\calO}}}
\nc{\tS}{{\tilde{S}}}
\nc{\spe}{{\on{sp}}}
\nc{\loc}{{\on{loc}}}
\nc{\Sym}{{\on{Sym}}}
\nc{\Cone}{{\on{C}}}
\nc{\syn}{{\on{syn}}}
\nc{\reg}{{\on{reg}}}
\nc{\colim}{{\on{colim}}}
\nc{\Norm}{{\on{N}}}

\nc{\nscrR}{{\mathscr{R}^{\on{nr}}}}

\nc{\GL}{{\on{GL}}}
\nc{\U}{{\on{U}}}
\nc{\Gl}{\on{Gl}} 
\nc{\GSp}{{\on{GSp}}}
\nc{\gl}{{\frakg\frakl}}
\nc{\SL}{{\on{SL}}} 
\nc{\SU}{{\on{SU}}} 
\nc{\SO}{{\on{SO}}}
\nc{\PGL}{{\on{PGL}}}

\nc{\Conv}{{\on{Conv}}}
\nc{\Rep}{{\on{Rep}}}
\nc{\Dom}{{\on{Dom}}}
\nc{\red}{{\on{red}}}
\nc{\act}{{\on{act}}}
\nc{\nr}{{\on{nr}}}
\nc{\ctf}{{\on{ctf}}}

\nc{\str}{{\on{-}}} 
\nc{\os}{{\bar{s}}}
\nc{\oeta}{{\bar{\eta}}}

\nc{\hookto}{\hookrightarrow}
\nc{\longto}{\longrightarrow}
\nc{\leftto}{\leftarrow}
\nc{\onto}{\twoheadrightarrow}
\nc{\lonto}{\twoheadleftarrow}

\nc{\uG}{{\underline{G}}}
\nc{\uA}{{\underline{A}}}
\nc{\uS}{{\underline{S}}}
\nc{\uT}{{\underline{T}}}
\nc{\uM}{{\underline{M}}}
\nc{\uP}{{\underline{P}}}
\nc{\uB}{{\underline{B}}}
\nc{\uN}{{\underline{N}}}

\nc{\ucG}{{\underline{\calG}}}
\nc{\ucA}{{\underline{\calA}}}
\nc{\ucS}{{\underline{\calS}}}
\nc{\ucT}{{\underline{\calT}}}
\nc{\ucalM}{{\underline{\calM}}}
\nc{\ucP}{{\underline{\calP}}}
\nc{\ucalN}{{\underline{\calN}}}

\nc{\bF}{{\breve{F}}}

\nc{\oFl}{{\overline{\Fl}}} 
\nc{\bU}{{\overline{U}}}
\nc{\tGr}{{\tilde{\Gr}}}
\nc{\cGr}{\calG\! r}
\nc{\oGr}{\overline{\on{Gr}}} 
\nc{\ocGr}{\overline{\calG\! r}}
\nc{\co}{{\colon}}
\nc{\sch}[1]{(Sch/{#1})}
\nc{\HypLoc}[1]{HypLoc({#1})}

\nc{\ohtimes}{\stackrel{!}{\otimes}}
\nc{\boxtilde}{\widetilde{\boxtimes}}
\nc{\vstar}{{\varhexstar}}

\nc{\Div}{\on{Div}}
\nc{\Sht}{\on{Sht}}
\nc{\Frob}{\on{Frob}}

\nc{\x}{\times}
\nc{\bsl}{\backslash}
\nc{\algQl}{{\bar{\bbQ}_\ell}}
\nc{\sF}{{\bar{F}}}
\nc{\nF}{{\breve{F}}}
\nc{\nW}{{W^{\on{nr}}}}
\nc{\sk}{{\bar{k}}}
\nc{\cont}{\on{c}}
\nc{\Supp}{\on{Supp}}
\nc{\blt}{\bullet}  
\nc{\dom}{\on{dom}}
\nc{\scon}{{\on{sc}}} 
\nc{\Affine}{\on{Aff}} 
\nc{\nscrA}{\mathscr{A}^{\on{nr}}} 
\nc{\nfraka}{{\bbf^{\on{nr}}}}
\nc{\ran}{{\rangle}}
\nc{\lan}{{\langle}}
\nc{\bk}{{\bar{k}}}
\nc{\tF}{{\tilde{F}}}
\nc{\sS}{{\bar{S}}}
\nc{\LG}{{^\text{L}\hspace{-0.04cm}G}}
\nc{\LL}{{^\text{L}\hspace{-0.07cm}L}}
\nc{\et}{{\text{\rm \'et}}}
\nc{\inv}{{\on{inv}}}
\nc{\Hecke}{{\on{Hecke}}}
\nc{\Isom}{{\on{Isom}}}
\nc{\oSht}{{\overline{\on{Sht}}}}
\nc{\AIJ}{{\calO_X[{\scriptstyle{\calI\over \calJ}}]}}
\nc{\Proj}{{\on{Proj}}}
\nc{\Bl}{{\on{Bl}}}

\nc{\Pos}{{\on{Pos}}}
\nc{\Sets}{{\on{Sets}}}
\nc{\AffSch}{{\on{AffSch}}}
\nc{\Groups}{{\on{Groups}}}
\nc{\Gpds}{{\on{Groupoids}}}
\nc{\Sch}{{\on{Sch}}}
\nc{\fl}{{\on{flat}}}

\nc{\pot}[1]{ [\hspace{-0,5mm}[ {#1} ]\hspace{-0,5mm}] }
\nc{\rpot}[1]{ (\hspace{-0,7mm}( {#1} )\hspace{-0,7mm}) }

\nc{\defined}{\hspace{0.1cm}\stackrel{\text{\tiny \rm def}}{=}\hspace{0.1cm}}
\begin{document}
\mainmatter              
\title{Formalizing All Indexed Mathematics as a Benchmark for General Reasoning: Implementing Dilatations of Categories}
\titlerunning{Formalizing All Indexed Mathematics as a Benchmark}  
%
\author{Arnaud Mayeux}
\authorrunning{A. Mayeux} 
%

%
\institute{University of Wisconsin-Madison, Madison, USA\\
\email{mayeux@wisc.edu}}

\maketitle
\begin{abstract}              
Formal rigor distinguishes mathematics from other disciplines, in the sense that mathematical statements are derived from explicit axioms by logically verifiable steps. Interactive theorem provers support this by expressing definitions, theorems, and proofs in a fully formal language and verifying them mechanically.
We consider the benchmark problem of formalizing all published mathematics as a machine-verifiable and continuously updated corpus of mathematical knowledge. This viewpoint treats mathematics as a structured database of interdependent results and raises questions about scalability and organization of large formal libraries.
As a case study, we present an ongoing formalization in categorical algebra, namely, dilatations of categories, extending classical localizations and illustrating what such an implementation looks like in practice.
\keywords{Formalizing all published mathematics, Benchmark, Theorem provers, Database, Dilatations of categories}
\end{abstract}
\section{Introduction}

This study is situated in the context of the growing interaction between mathematics and computer-assisted reasoning systems. Recent developments in interactive theorem provers have made the formal expression of definitions, theorems, and proofs in a fully formal language, and their mechanical verification, both possible and increasingly accessible in practice.

In parallel, libraries such as Mathlib in the Lean theorem prover system show that substantial parts of modern mathematics are already represented in machine-checkable form.

We motivate the benchmark problem of formalizing all indexed mathematics in a machine-verifiable system and maintaining it as a continuously updated formal corpus. We argue that, at a theoretical level, the program is feasible in principle, and we highlight organizational and infrastructural challenges.

Within this framework, we present an ongoing case study in the formalization of categorical constructions, focusing on dilatations of categories, which extend the classical theory of localization. This serves as a concrete example of how new mathematical structures can be developed within an existing formal library.

\section{Mathematics and Rigor} \label{sec-math-rigor-1}
In principle, by the very definition of mathematics, mathematics gives us results derived with absolute rigor, based on axioms. This absolute certainty makes mathematics special among sciences. In applications to other sciences, mathematics does not always guarantee truth or practical relevance; other sciences require empirical evidence, interpretations, and contextual understanding. As a pillar, mathematics must remain fully rigorous. However, we have reached a time when mathematics is vast, and sometimes errors arise due to its expansion and human imperfection in checking details. On the other hand, as anticipated a long time ago (cf. e.g. \cite{NSS56,Ha61}), we now have machines able to compute the validity of a fully formalized proof\footnote{Theorem proving began with the Logic Theorist \cite{NSS56}, the first AI program to prove mathematical theorems. Interactive systems like LCF (Milner, 1972) \cite{Mi72,Mi84}, Automath (de Bruijn, 1970) \cite{Br70}, HOL (Gordon, 1981) \cite{Go87}, Coq (Coquand, Huet, 1988) \cite{CH88}, and Isabelle (Paulson, 1990) \cite{Pa90} integrated computing and formal logic for verification. Lean and Mathlib aim to scale formalization to the broader mathematical literature \cite{dM15,dMU21,Mathlib}.}. Theorem provers like Lean can renew the special status of mathematics regarding rigor. Theorem provers transform the verification of rigor into formal computations performed by highly reliable physical systems.
As explained, we now have a way to check and verify mathematics mechanically. Currently, it is humans who must communicate with the machine and explain all the details. There is already an AI-generative tool in Lean that, in principle, can fill in the proof by itself, but, as of today, AI systems get stuck very quickly when formalizing advanced math. For instance, they get stuck on the very first lines of a random published article. Most mathematical results build on earlier ones, and formalizing a result also requires formalizing what it relies on. We refer to \cite{AH14,dM15,Bu19,dMU21,Av22,Av24,Bu24,Mas24,ACMT25} for recent surveys on Lean and theorem provers written by researchers who have long been active in this field.

\section{Formalizing all Mathematics as a Benchmark} \label{sec-bench-2}

We aim to develop general AI systems capable of supporting a wide range of human reasoning routines. The previous sentence is slightly informal, and it is natural to list some of these routines for clarity. Until very recently, almost all existing mathematics had been produced and verified by humans. Any system claiming to be general must be able to perform at least all basic tasks that humans can, including verifying advanced mathematical results. Thus, a general intelligent system has to contain all indexed mathematics in a formalized way.  

This suggests the following benchmark: formalize all published mathematics and update it in real-time. In some form, such ideas can be traced back to \cite{Boy94,KoRa16}.

As of 2025, the main database for mathematics, MathSciNet, lists 4.5 millions publications. 
Whatever the estimate for the average size required to formalize a mathematical publication, these data could easily be stored in a relatively small data center. Once this is done (even partly), the integration of these data with artificial reasoning and generative methods \cite{MSH16,Xu25} will provide a very powerful method for mathematical research. Without full formalization of all indexed papers, current AI is not reliable enough for rigorous original mathematical research at a high level. Fully formalized mathematics is the key for developing serious AI-powered tool for mathematical research.

\section{Organizing Formalization} \label{sec-org-3}

Formalizing all mathematics is very important. Improving the tool we use is important: one can only acknowledge smoother version and more user friendly one. However, once we have a theorem prover which in principle can support all of mathematics, the major difference will be if actually all of mathematics is implemented or not. This would require the involvement of mathematicians themselves, not only computer scientists\footnote{We already have hundreds of mathematicians involved with Lean and Mathlib, and the number of participants continues to grow. However, as of today, this still represents only a small proportion of mathematicians.}.
 So it is crucial to make mathematicians themselves want to participate in this enterprise. Our main point here is that compatibility between versions of Lean is crucial: more mathematicians will be willing to invest time in the language if they are confident that it will not become obsolete or incompatible within a few years. The following is a list of comments on the formalization of mathematics. We make no claim of originality, as these questions arise naturally in this context.

\begin{enumerate}
\item As a basic principle, new formalization depends on what is already formalized, because theorems and definitions rely on previous theorems and definitions.

\item Each contributor will inherently influence what can be formalized next.

\item One would want a database for all formalized mathematics. Today, Mathlib (\url{http://leanprover-community.github.io/mathlib4_docs/}) resembles Bourbaki more than MathSciNet or zbMATH Open: a useful polished foundational selective library rather than a comprehensive catalog of mathematics. Reservoir (\url{http://reservoir.lean-lang.org/}) points toward that broader vision by indexing Lean packages and repositories, tracking dependencies and compatibility, and providing searchable metadata and documentation. At present, Reservoir operates mainly at the level of projects, providing infrastructure for archiving and indexing formalization projects. To our knowledge, MathSciNet and Reservoir are not currently integrated, and there does not yet exist a unified established infrastructure connecting large-scale mathematical publication indexing with large-scale formalized mathematics repositories.

\item In case new versions of Lean appear, it is important to have a way to translate older formalizations into the new version to keep all formalized material compatible. In our opinion, this is fundamental to convince everybody that formalizing some mathematics is not time wasting. 

\item When AI-assisted auto-formalization become better and better, we will have to focus carefully on the concordance on what AI produces and what the original paper actually said. This verification also applies to current human-produced formalizations. The database should include a section indicating the corresponding mathematical references and the formalization journal that validated each concordance. Including a user commentary section could be useful too. 

\item A decisive step will occur when the speed of formalization surpasses the production of new unformalized mathematical knowledge. As of today, the volume of new mathematical publications is still much higher than the volume of formalization projects. 

\item When a significant part of mathematics is formalized and a significant portion of new results relies on formalized mathematics, some pure math journals may start adding a ‘formalized’ label to publications. Over time, by verified mathematics, new generations of mathematicians may mean by default what is formalized in such a system.

\end{enumerate}

\noindent In the rest of this study, we will illustrate the above discussion on the example of dilatations of categories. Dilatation of categories is a very general and recently published construction in category theory. This construction relies on localization of categories. Localization of categories, introduced by Gabriel-Zisman \cite{GZ67}, is now viewed as foundational and it has been formalized in Lean 4 and included in Mathlib. 

Looking at the material of the formalization of localization of categories, we observed that it is accessible to formalize dilatations of categories. Therefore, we decided to do this formalization. 
 We mostly produced this formalization ourselves, as current AI systems were not able to correctly formalize this result. In this case, we, therefore, had to guide the formalization process manually. Again, the more the system is trained on various kinds of formalized mathematics, the more it will be able to perform later.

\section{Categorical Algebra in Lean} \label{sec-cat-lean-4}
\subsection{What are Dilatations?} \label{sub-whatis1}
If one has a structure with a composition law (e.g., multiplication of numbers, or composition of arrows), which happens very often in mathematics, one is familiar with the concept of fraction. A fraction is the quotient of a numerator by a denominator. By definition, the fraction composed appropriately with the denominator of the fraction equals the numerator. Dilatations form a very general process that adds fractions to a structure in a universal way \cite{Ma25}.

This refines localization \cite{GZ67}, a well-known construction which is already formalized and part of Mathlib. We use much of the material on localization in this work.

We now fix a category  $\calC$ with objects $Ob\calC $ and morphisms $Mor \calC$. Let $\Sigma $ be a collection of morphisms of $\calC$. Let us recall some properties of the localization $\calC[\Sigma ^{-1}]$. We have a functor $L:\calC \to \calC[\Sigma^{-1}]$. The objects of the localization $\calC [\Sigma ^{-1}]$ coincide with the objects of $\calC$, i.e. $L$ is the identity on objects. Given a morphism $d$ in $\Sigma $, $L(d)$ is an isomorphism in $\calC[\Sigma ^{-1}]$. If $F: \calC \to \calD$ is another functor such that $F(d) $ is an isomorphism for all $d$ in $\Sigma$, then $F$ factors through $L$.
 Now assume that for any $d$ in $\Sigma$ we have a sieve $N_d$, in $\calC$ over the codomain of $d$. The dilatation process will provide a category $\calC '$ and a functor $\Theta : \calC \to \calC'$. The objects of $ \calC'$ will also coincide with the objects of $\calC$. For any $d \in \Sigma$ and any $n \in N_d$, there exists a unique arrow $b$ in $\calC '$ such that the diagram: 
\[ \begin{tikzcd}
X \ar[rr, "{\Theta (n)}"] \ar[rd, "{b}", dotted] & & Y \\
& Z \ar[ru, "{\Theta (d)}" ] &
\end{tikzcd}\] commutes. The element $b$ is thought of as a non-commutative fraction $d \backslash n = d^{-1} \circ n $.
Dilatations also satisfy a natural universal property \cite{Ma25}, namely $\Theta $ is universal among all functors $F : \calC \to \calD$ such that: 
\begin{enumerate}
\item The sieve generated by $F(N_d) $ is included in the sieve generated by $F(d)$, for all morphisms $d$ in the collection $\Sigma$,
\item The localization $\calD \to \calD [\Sigma ^{-1}]$ is faithful. 
\end{enumerate}

\noindent In the following, we discuss our ongoing formalization of dilatations of categories in Lean. The repository is available at: 
https://github.com/rndmx/DilCat. 

All the material presented here compile correctly. The final version may differ slightly in some parts of the code.

\subsection{Sieves} \label{sub-sieve2}

 Recall that if $X\xrightarrow{a}Y$ belongs to $Mor \calC$, $dom(a):= X$ is called the domain and $cod(a):=Y$ the codomain (note that $dom$ and $cod$ also make sense for directed graphs). Recall that a sieve in $\calC$ with codomain $X$ is a collection of morphisms stable by precompositions and whose codomains are $X$.
This notion has been implemented in Lean in: 

\lstinline{Mathlib.CategoryTheory.Sites.Sieves}.

\begin{lstlisting}
structure Sieve {C : Type u₁} [Category.{v₁} C] (X : C) where

  arrows : Presieve X

  downward_closed : ∀ {Y Z f} (_ : arrows f) (g : Z ⟶ Y), arrows (g ≫ f)
\end{lstlisting}

\noindent A presieve is a set of arrow with fixed codomain. 

\begin{lstlisting}

def Presieve (X : C) :=
  ∀ ⦃Y⦄, Set (Y ⟶ X)
\end{lstlisting}

\subsection{Directed graphs and localizations}
\label{subsectgraph3}
 In this section, we discuss localizations of categories.  The content of this subsection was introduced in \cite{GZ67}. Here we follow the treatment in \cite{Ma25} and the formalization in:
 
  \lstinline{Mathlib.CategoryTheory.Localization.Construction}.
  Let $\Sigma $ be a collection of morphisms of $\calC$. In the formalization $W$ corresponds to $\Sigma$ in \cite{Ma25}.

\defi \label{graphdef} 
Let $\calG $ be the oriented graph (quiver) defined as follows: The vertices of $\calG$ are equal to the objects of $\calC$. The directed lines of $\calG$ are made of: \begin{enumerate} \item For each morphism $a$ of $\calC$, a directed line $dom(a) \xrightarrow{a} cod(a)$ of $\calG $,
\item For each morphism $d$ in $\Sigma $ a directed line $cod(d) \xrightarrow{l_d} dom(d)$ of $\calG$ [in particular $dom (l_d) = cod(d) $ and $cod (l_d) = dom (d)$].
\end{enumerate}
\xdefi

\begin{lstlisting}
universe uC' uD' uC uD
variable {C : Type uC} [Category.{uC'} C] (W : MorphismProperty C) {D : Type uD} [Category.{uD'} D]

namespace Localization

namespace Construction


structure LocQuiver (W : MorphismProperty C) where
  
  obj : C

instance : Quiver (LocQuiver W) where Hom A B := (A.obj ⟶ B.obj) ⊕ { f : B.obj ⟶ A.obj // W f }
\end{lstlisting}

\defi
A $\Sigma$-sequence of directed lines in $\calG$ is a sequence, finite and possibly empty, of directed lines $x_1, x_2 , \ldots , x_n$ of $\calG$ such that  $cod(x_i)= dom (x_{i+1})$ ($n\geq0$ is an integer). By convention, an empty sequence is just an element in the collection $Ob \calC$. For a non-empty sequence $s=(x_1 , \ldots , x_n )$, we define the domain as $dom(s )= dom (x_1) $ and the codomain as $cod (s) = cod (x_n)$. If a sequence $s$ is empty and given by an object $X$, then we put $dom(s)=cod(s)=X$. Note that $\Sigma$-sequences of directed lines with compatible domains and codomains can be composed (by convention, composing a sequence $s$ with an empty sequence $e$ gives $s$).
\xdefi

\defi  \label{sequencedefilict}
We say that two  $\Sigma$-sequences of directed lines $s$ and $s'$ in $\calG$ are equivalent if $dom(s)=dom(s'),$ $ cod(s)=cod(s')$ and if one can be obtained from the other by a chain of elementary equivalences of the following types:
\begin{enumerate}
\item A sequence  $x_1, \ldots , x_n $ such that  $x_i , x_{i+1}$  are equal to $a,a' \in Mor \calC$ and $cod(a)=dom(a')$, for some $1 \leq i , i+1 \leq n$, is equivalent to the sequence $x_1 , \ldots , x_{i-1} , x' , x_{i+2}, \ldots , x_{n}$ where $x' $ is the composition of $a $ and $a'$ in $\calC$,
\item A sequence $x_1 , \ldots , x_n $ such that $x_i , x_{i+1} $ are equal to $ l_d , d$ with $d \in \Sigma$, for some $1 \leq i , i+1 \leq n$, is equivalent to the sequence 
$x_1, \ldots , x_{i-1} , x_{i+2}, \ldots , x_n $,
\item A sequence $x_1 , \ldots , x_n$ such that $x_i , x_{i+1}$ are equal to $d,l_d$ with $d \in \Sigma $, for some $1 \leq i , i+1 \leq n$, is equivalent to the sequence 
$x_1, \ldots , x_{i-1} , x_{i+2}, \ldots , x_n$,
\item for any object $X$, the sequence $Id_X$ is equivalent to the empty sequence at $X$.
\end{enumerate} 
\noindent In other words and informally, two $\Sigma$-sequences of directed lines are equivalent if one can be obtained from the other  by the operations consisting in exchanging parts of sequences as follows: \begin{align*}
  X \xrightarrow{a} Y \xrightarrow{a'} Z  & \leftrightsquigarrow  X \xrightarrow{a' \circ a } Z \\    Y \xrightarrow{l_d } Z \xrightarrow{d} Y & \leftrightsquigarrow  Y  \text{ (the empty sequence at $Y$)} \\   Z \xrightarrow{d} Y \xrightarrow{l_d} Z & \leftrightsquigarrow   Z  \text{ (the empty sequence at $Z$)} \\ X \xrightarrow{Id_X} X & \leftrightsquigarrow X  \text{ (the empty sequence at $X$)}.
\end{align*} The equivalence class of a sequence $s$ is denoted by $[s]$.
Note that equivalence classes of $\Sigma$-sequences of directed lines with compatible domains and codomains can be composed associatively. 
\xdefi

\defi A $\Sigma$-fraction is an equivalence class of $\Sigma$-sequence.

\xdefi

\begin{lstlisting}

def ιPaths (X : C) : Paths (LocQuiver W) :=
  ⟨X⟩


@[simp]
def ψ₁ {X Y : C} (f : X ⟶ Y) : ιPaths W X ⟶ ιPaths W Y := (Paths.of _).map (Sum.inl f)

@[simp]
def ψ₂ {X Y : C} (w : X ⟶ Y) (hw : W w) : ιPaths W Y ⟶ ιPaths W X :=
  (Paths.of _).map (Sum.inr ⟨w, hw⟩)


inductive relations : HomRel (Paths (LocQuiver W))
  | id (X : C) : relations (ψ₁ W (𝟙 X)) (𝟙 _)
  | comp {X Y Z : C} (f : X ⟶ Y) (g : Y ⟶ Z) : relations (ψ₁ W (f ≫ g)) (ψ₁ W f ≫ ψ₁ W g)
  | Winv₁ {X Y : C} (w : X ⟶ Y) (hw : W w) : relations (ψ₁ W w ≫ ψ₂ W w hw) (𝟙 _)
  | Winv₂ {X Y : C} (w : X ⟶ Y) (hw : W w) : relations (ψ₂ W w hw ≫ ψ₁ W w) (𝟙 _)
\end{lstlisting}

\defi The localization of $\calC $ relatively to $\Sigma$ is the category $\calC [\Sigma^{-1}]$ whose objects are the objects of $\calC$ and whose morphisms are $\Sigma$-fractions. We have a canonical functor $L:\calC \to \calC[\Sigma^{-1}]$.
\xdefi 

\begin{lstlisting}
def Localization :=
  CategoryTheory.Quotient (Localization.Construction.relations W)

instance : Category (Localization W) := by
  dsimp only [Localization]
  infer_instance

def Q : C ⥤ W.Localization where
  obj X := (Quotient.functor _).obj ((Paths.of _).obj ⟨X⟩)
  map f := (Quotient.functor _).map (ψ₁ W f)
  map_id X := Quotient.sound _ (relations.id X)
  map_comp f g := Quotient.sound _ (relations.comp f g)
\end{lstlisting}

\prop[Universal Property of Localization]
Let $F: \calC \to \calD$ be a functor such that $F(d)$ is invertible for all $d $ in $\Sigma$. Then there exists a unique functor $F': \calC[\Sigma ^{-1}] \to \calD$ such that  $F= F' \circ L$.
\xprop

\begin{lstlisting}
@[simps!]
def lift : W.Localization ⥤ D :=
  Quotient.lift (relations W) (liftToPathCategory G hG)
    (by
      rintro ⟨X⟩ ⟨Y⟩ f₁ f₂ r
      rcases r with ⟨⟩ <;> all_goals aesop)
      
theorem uniq (G₁ G₂ : W.Localization ⥤ D) (h : W.Q ⋙ G₁ = W.Q ⋙ G₂) : G₁ = G₂
\end{lstlisting}

\subsection{Implementing Dilatations of Categories: Centers} 
\label{sub-sectdef4}

Let $\calC$ be a category. 

\defi
A center in $\calC$ is a collection  $\{[N_i, d_i ]\}_{i \in I }$ of pairs $[N_i , d_i]$, indexed by a collection $I$ and such that, for all $i$ in $I$, $d_i$ is a morphism of $
\calC$ and $N_i$ is a sieve over $cod(d_i)$.
\xdefi

\begin{lstlisting}
structure Center (C : Type u) [Category.{v} C] where
  I : Type u
  (nonempty : Nonempty I)
  dom : I → C
  cod : I → C
  mor : ∀ i : I, dom i ⟶ cod i
  N   : ∀ i : I, Sieve (C := C) (cod i)
\end{lstlisting}

\noindent We now introduce the localized category and the quiver associated to the morphisms \lstinline{mor} applying the material in Mathlib.

\begin{lstlisting}
def IsCenterMor (f : Σ X Y : C, X ⟶ Y) : Prop :=
  ∃ i : Z.I, f = ⟨Z.dom i, Z.cod i, Z.mor i⟩

def CenterMorphismProperty : MorphismProperty C := fun X Y f => IsCenterMor Z ⟨X, Y, f⟩

def CenterLocalization : Type u := (CenterMorphismProperty Z).Localization

def LocalizationFunctor : C ⥤ (CenterMorphismProperty Z).Localization := (CenterMorphismProperty Z).Q

def Quiv := LocQuiver (CenterMorphismProperty Z)
\end{lstlisting}
 
\subsection{Implementing Dilatations of Categories: Prescribed Fractions}

We now fix a center $\{[N_i, d_i ]\}_{i \in I }$ in $\calC$ (we sometimes use the notation $N_{d_i}$ to denote $N_i$). 

Put $\Sigma =\{d_i\}_{i \in I}$.

\defi \label{fractiondef} A $\{[N_i, d_i ]\}_{i \in I }$-fraction is a $\Sigma$-fraction such that a representative can be written as:
\begin{center}$ X_1 \xrightarrow{n_1} Y_1 \xrightarrow{l_{d_{i_1}}} X_2 \xrightarrow{n_2} Y_2 \xrightarrow{l_{d_{i_2}}} X_3 \ldots X_{k} \xrightarrow{n_k} Y_k \xrightarrow{l_{d_{i_k}}} X_{k+1} \xrightarrow{a} X_{k+2}$\end{center}  with $a \in Mor \calC$, $k \geq 0$ and ${i_j} \in I$, $n_i \in N_{i_j} $ for all $j \in \{1 , \ldots , k \}.$
\xdefi

\begin{lstlisting}
structure Fraction :=
  (k : ℕ)
  (Xs : ℕ → C) 
  (Ys : ℕ → C)
  (i  : ℕ → Z.I) 
  (n_orig : Π (j : {j // j < k}), Xs j ⟶ Ys j)
  (n_dom : ∀ j : {j // j < k}, Z.dom (i j) = Xs (j+1))
  (n_cod : ∀ j : {j // j < k}, Z.cod (i j) = Ys j)
  n : Π (j : {j // j < k}), Xs j ⟶ Z.cod (i j) :=
   λ j => n_orig j ≫ eqToHom (n_cod j).symm
  (n_in_N : ∀ j : {j // j < k }, Z.N (i j) (n j))
  (a : Xs (k) ⟶ Xs (k+1))

def inv_in_path (F: Fraction Z)  (j : {j // j < (F.k )}):
  ιPaths (CenterMorphismProperty Z) (Z.cod (F.i j)) ⟶
  ιPaths (CenterMorphismProperty Z) (Z.dom (F.i j)) :=
  Localization.Construction.ψ₂ (CenterMorphismProperty Z)
  (Z.mor (F.i j)) ⟨F.i ↑j, rfl⟩

def fraction_in_path_single (F : Fraction Z) (j : {j // j < F.k }) :
  ιPaths (CenterMorphismProperty Z) (F.Xs j) ⟶
  ιPaths (CenterMorphismProperty Z) (F.Xs (j+1)) :=
  Localization.Construction.ψ₁ (CenterMorphismProperty Z) (F.n j) ≫
    inv_in_path Z F j ≫
      eqToHom (congrArg (ιPaths (CenterMorphismProperty Z)) (F.n_dom j))

def fraction_in_path_seq (F : Fraction Z) :
  Π (n : {n // n < F.k }),
    ιPaths (CenterMorphismProperty Z) (F.Xs 0) ⟶ ιPaths (CenterMorphismProperty Z) (F.Xs (n+1))
| ⟨0, h⟩ => fraction_in_path_single Z F ⟨0, h⟩
| ⟨n+1, h⟩ =>
  let prev : {j // j < F.k } := ⟨n, Nat.lt_of_succ_lt h⟩
  fraction_in_path_seq F prev ≫ fraction_in_path_single Z F ⟨n+1, h⟩

def fraction_in_path_last (F : Fraction Z) :
    ιPaths (CenterMorphismProperty Z) (F.Xs 0) ⟶ ιPaths (CenterMorphismProperty Z) (F.Xs (F.k)):=
if h : F.k = 0 then
  eqToHom (congrArg (ιPaths (CenterMorphismProperty Z)) (by rw [h]))  
else
  let x := F.k - 1
  have hx : F.k-1 < F.k := Nat.sub_lt (Nat.pos_of_ne_zero h) (Nat.succ_pos _)
  fraction_in_path_seq Z F ⟨F.k - 1, hx⟩ ≫
  eqToHom (congrArg (ιPaths (CenterMorphismProperty Z)) (congrArg (F.Xs) (Nat.succ_pred_eq_of_pos (Nat.pos_of_ne_zero h))))

def fraction_in_path_full (F : Fraction Z) :
    ιPaths (CenterMorphismProperty Z) (F.Xs 0) ⟶ ιPaths (CenterMorphismProperty Z) (F.Xs (F.k+1)):=
    fraction_in_path_last Z F  ≫
      Localization.Construction.ψ₁ (CenterMorphismProperty Z) (F.a)

def fraction_in_loc_full (F : Fraction Z)  :
objEquiv (CenterMorphismProperty Z) (F.Xs 0) ⟶
objEquiv (CenterMorphismProperty Z) (F.Xs (F.k+1)) :=
 (CategoryTheory.Quotient.functor (relations (CenterMorphismProperty Z))).map (fraction_in_path_full Z F)
\end{lstlisting}

\noindent The dilatation of $\calC$ with center $\{[N_i, d_i ]\}_{i \in I }$ is the category $\calC'$ defined as follows. The objects of $\calC'$ are equal to the objects of $\calC$. If $X $ and $Y$ are objects in $\calC '$, morphisms between $X $ and $Y$ are given by $\{[N_i, d_i ]\}_{i \in I }$-fractions (cf. Definition \ref{fractiondef}) with domain $X$ and codomain $Y$. 

We plan to formalize the following theorem as part of the continuing development of this work in progress.

\theo\label{uni} (Universal property)  Let $F: \calC \to \calD$ be a $\Sigma$-regular functor (i.e. $\calD \to \calD [\Sigma^{-1}]$ is faithful, cf. \cite{Ma25}) such that
for any $i $ in $I$,  we have: \[S^{\calD}_{F(N_i)} \subset S^{\calD}_{F(d_i)} .\]
Then there is a unique functor $F' : \calC ' \to \calD$, such that the triangle of functors: \[
\begin{tikzcd}
\calC \ar[rr, "F"] \ar[rd, "\Theta "] & & \calD \\
& \calC ' \ar[ru, " F' " ] &
\end{tikzcd}  \] commutes. 
\xtheo

\subsection{Limitations}
\begin{enumerate}
\item The proposed benchmark of formalizing all indexed mathematics has limitations, as it depends on the clarity, consistency, and completeness of existing mathematical literature, and some published results may be incorrect, underspecified, or ambiguous, and may therefore require correction, clarification, or reformulation before formalization.

\item Requiring all work to be fully formalized may limit the creativity of some visionary research. However, this is not a major problem, as it remains sufficient to distinguish strict, formal mathematics from more exploratory approaches in the field. It is still possible to publish visionary work that is not fully formalizable.

\item A key question is whether full formalization of the mathematical literature is feasible at scale. The global mathematical corpus, as indexed in major repositories such as AMS-related databases, contains approximately $4.5 \times 10^6$ publications.
\end{enumerate}
\noindent An estimate of the formalization effort ranges from $0.5$ to $50$ person-months per paper. For the corpus of $4.5 \times 10^6$ publications, this yields:

\begin{align*}
4.5 \times 10^6 \times 0.5 &= 2.25 \times 10^6 && \text{person-months} \quad (\approx 1.88 \times 10^5 \ \text{person-years}), \\
4.5 \times 10^6 \times 50 &= 2.25 \times 10^8 && \text{person-months} \quad (\approx 1.88 \times 10^7 \ \text{person-years}).
\end{align*}

\noindent Assuming partly AI-assisted formalization can provide efficiency gains between $5\times$ and $20\times$, the effective effort is reduced to approximately $9.4 \times 10^3$--$3.75 \times 10^6$ person-years.

From a data infrastructure perspective, the total corpus size is quite small compared to modern industrial systems \cite{BHR26}. Hyperscale datacenters routinely operate at the exabyte scale ($10^{18}$ bytes). 

Assuming a combined representation size (PDF, LaTeX, Lean formalization, and associated integrating data) of approximately 1--100 MB per publication, the full mathematical corpus of $4.5 \times 10^6$ papers is estimated to require between $4.5 \times 10^{12}$ bytes ($\approx$ 4.5 TB) and $4.5 \times 10^{14}$ bytes ($\approx$ 450 TB).

This places the problem in a regime where computational storage is not limiting; instead, the dominant difficulty lies in structural transformation of unformalized knowledge into a consistent machine-verifiable representation.

While significant coordination and sustained effort will be required, both technological trends and community momentum suggest that full formalization of the indexed mathematical literature is a realistic long-term objective. It is plausible that in the near future, autoformalization systems will be able to produce formalizations of some papers in seconds. For some papers, a human expert could then review and validate the autoformalized version in just a few minutes. This would dramatically reduce the effective effort.

\subsubsection{Declaration.}

The author declares that all materials reproduced from other sources are used with appropriate permissions and are duly cited. 
The author has no conflicts of interest that could affect the integrity or objectivity of this work. This section is independent and has no connections or endorsements with other contributions in this volume. All funding, institutional support, and contributions from individuals or organizations are gratefully acknowledged. 
The opinions expressed in this section are solely those of the author and do not necessarily reflect the views of affiliated institutions. Generative AI tools may have been used for language refinement, but the author retains full responsibility for the content.

\subsubsection{Acknowledgment.}

The author thanks the anonymous reviewers for their constructive critiques, comments, and supportive feedback, which improved the quality of the manuscript.

\end{document}